\documentclass[%
 preprint, 
 amsmath,amssymb,
 aps, physrev,
]{revtex4-2}

\usepackage{graphicx}
\usepackage{dcolumn}
\usepackage{bm}

\usepackage{xcolor}

\begin{document}


\title{\textbf{Electron Orbital Angular Momentum Polarization in Neutral Atoms} 
}%

\author{Hongtao Hu}
\author{Andrius Baltuška}%
\affiliation{%
 Technische Universität Wien, Photonics Institute,
 Gußhausstraße 27-29, Vienna A-1040, Austria
}%

\author{Sebastian Mai}
\affiliation{University of Vienna, Institute of Theoretical Chemistry, Faculty of Chemistry, 1090 Vienna, Austria}

\author{Peng Peng}
 \email{pengpeng@shanghaitech.edu.cn}
\affiliation{
 ShanghaiTech University, School of Physical Science and Technology, Shanghai 201210, Shanghai, China
}%
\affiliation{
ShanghaiTech University, Center for Transformative Science, Shanghai 201210, Shanghai, China
}%

\author{Xinhua Xie}
\email{xinhua.xie@psi.ch}
\affiliation{
SwissFEL, Paul Scherrer Institute, Villigen PSI 5232, Switzerland
}%


\date{\today}

\begin{abstract}

We demonstrate the polarization of electron orbital angular momentum (OAM) in neutral atoms by integrating the Zeeman effect with attosecond transient absorption spectroscopy (ATAS). Using density matrix simulations, we show that in a helium atom, the absorption probability asymmetry between 
$m_j = -1$ and $m_j = 1$ in the $1s2p$ state can be precisely controlled by adjusting the time delay between infrared (IR) and extreme ultraviolet (XUV) fields, the strength of an applied static magnetic field, as well as the angle between laser polarization and magnetic field direction.
This approach has significant implications across various fields, including quantum computing, quantum communication, and spintronics.
Moreover, it paves the way for advancements in applications such as manipulating chemical reactions control, tailoring the magnetic properties of matter, and enabling novel laser emissions.

\end{abstract}

\maketitle

\vspace{-0.4cm}
\section{\label{sec:introduction}Introduction}

\vspace{-0.25cm}
Orbital angular momentum (OAM) plays a fundamental role in atomic and molecular structures, as well as in the broader field of quantum mechanics \cite{Heisenberg1925,Schroedinger1926,Sakurai1994}.  
Beyond its theoretical importance, orbital angular momentum underpins key applications in quantum computing \cite{Daniel1998}, quantum communication \cite{Hensen2015}, and spintronics  \cite{Hirohata2020}, where precise control of quantum states is essential.  
As modern technologies increasingly rely on tailored quantum state manipulation, developing methods to control electron OAM in neutral atomic systems has become a critical area of research.

While previous studies \cite{Dellweg2017,Eckart2018,Hartung2016,Liu2018,Stammer2023} have primarily focused on spin angular momentum (SAM) control, the manipulation of OAM states in neutral atoms remains relatively unexplored. 
Notably, research by Hartung et al. \cite{Hartung2016} successfully demonstrated electron spin polarization in Xenon ionization processes, but practical applications often require quantum systems to remain in their stable, neutral states \cite{Wolfgang2018}. 
Moreover, OAM's role extends beyond spin dynamics, influencing the spatial structure of electronic wavefunctions that govern complex phenomena such as molecular interactions, chemical reactivity, and magneto-optical properties. 
Therefore, exploring new ways to selectively control OAM states in neutral atoms is both scientifically intriguing and technologically relevant.

In this work, we achieve the polarization of the electron OAM in neutral atoms by combining the Zeeman effect \cite{Zeeman1897,Crutcher2019} with the attosecond transient absorption spectroscopy (ATAS) \cite{Huillier1988,Corkum1993,Krausz2001,Goulie2010,Leone2016,Borrego2022,Pfeifer2016}.
While ATAS has proven a powerful tool for probing ultrafast dynamics in neutral systems \cite{Peng2022,Peng2019,Santra2011}, its integration with the Zeeman effect has not been reported to date. 
Here, we demonstrate such integration for the first time, representing a significant advancement in the field since it enables attosecond and femtosecond pulses to directly access the electronic sub-states in atoms.
The concept is that a static magnetic field facilitates the Zeeman splitting of orbitals with distinct orbital-angular momentum. Subsequently, XUV and IR pulses are employed to excite the ground-state atoms to the splitting sub-states via electric dipole transitions. 
By manipulating the time delay between XUV and IR fields, the strength of the static magnetic field, and the angle between the laser polarization and magnetic field direction, the absorption probabilities of sub-states with different orbital quantum numbers can be controlled.
Specifically, taking He as an example, the asymmetrical factor calculated from two orbitals with $m_j=1$ and $-1$ in the $1s2p (^1P_1)$ state can be precisely controlled within a range of $-80\%$ to $40\%$ under realistic parameters.
It is necessary to emphasize that this approach is not restricted to OAM control or confined to atomic systems.
It holds promise for manipulating spin angular momentum and can be adapted for broader applications in molecular systems, significantly expanding its potential utility.

\section{Method} \label{sect:method}

\vspace{-0.25cm}

\begin{figure}[hbt]
    \centering
    \includegraphics[width=0.95\textwidth]{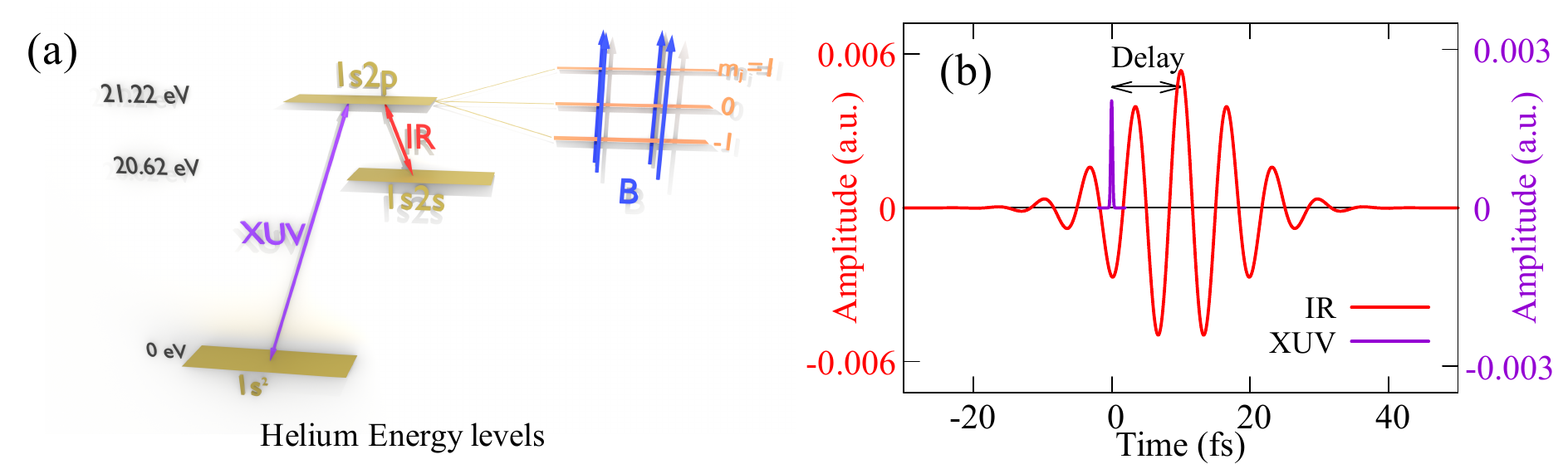}
    \caption { (a) Schematic diagram of helium energy levels excited by XUV and IR fields. In the presence of a magnetic field, the $1s2p (^1P_1)$ state will split into three states due to the Zeeman effect. (b) Temporal profiles of the XUV and IR fields with the full width half maximum of $0.2 fs$ and $20 fs$, respectively. A positive delay indicates that the IR field precedes the XUV field.  } 
    \label{fig:1_laser}
\end{figure} 


Our simulations employ the density matrix method \cite{Heisenberg1925,BornJordan1925,BornHeisenJordan1926,Dirac1925}, a computationally efficient framework for simulating quantum dynamics, with the ability to focus on the most relevant quantum states of the system \cite{Goldberg2021,Blum2012,Ma2022,Haffner2008}. 
Using density matrix method, we successfully reproduce the attosecond transient absorption spectroscopy results reported in Ref. \citenum{Wu2016}, which were originally obtained through the time-dependent Schrödinger equation simulations.
The time evolution of the density matrix is governed by the Liouville-von Neumann equation \cite{Breuer2003}:
\begin{equation}
\label{eq:DMM}
\frac{d{\hat{\rho}}}{dt}=-\frac{i}{\hbar}[\hat{H},\hat{\rho} ],
\end{equation}
where $t$ represents time, $\hat{H}$ and $\hat{\rho}$ are the total Hamiltonian and density matrix, respectively.
$\hat{\rho}$ is defined as $|\psi\rangle\langle\psi|$, where $|\psi\rangle=\sum_ia_i|\varphi_i\rangle$ and $|\varphi_i\rangle$ are the eigenstates of the field-free system.
The system is initially in its ground state, represented by $\hat{\rho}_{11}(t=0)=1$ (with state~$1$ being the $1s^2$ state), while all other density matrix elements are zero: $\hat{\rho}_{mn}(t=0)$ for $(m,n)\neq(1,1)$.
This assumption is justified by the large energy gap between the $1s2p$ and $1s^2$ states ($21.22 eV$), which far exceeds the thermal energy of molecular motion at room temperature ($kT=0.026 eV$).
The time evolution of this equation is carried out with a time step of $0.1 a.u.$ to ensure numerical stability and accuracy.

The total Hamiltonian consists of three terms, $\hat{H}=\hat{H}_0+\hat{V}_e+\hat{V}_m$, calculated in the Cartesian basis (see Supplementary Material for more details).
The first term ($\hat{H}_0$) is the Hamiltonian in the absence of the external fields. 
Its diagonal elements correspond to the eigen energies of the field-free system, which are time independent, and all the non-diagonal elements are zero. 
Figure \ref{fig:1_laser} (a) illustrates the three lowest energy states of the singlet Helium atom. 
It has been confirmed that incorporating higher states, such as $1s3s$ or $1s3d$, has a negligible effect on the results.
The second term ($\hat{V}_e$) describes the interaction between the atom and the electric laser fields: $\hat{V}_e=\hat{d_x}\cdot\varepsilon_x(t)+\hat{d_y}\cdot\varepsilon_y(t)+\hat{d_z}\cdot\varepsilon_z(t)$. 
Here, $\hat{d}_{x,y,z}$ represents the dipole transition matrix, with transition value $\hat{d}_{x,y,z}(1s^2\rightarrow1s2p_{x,y,z})=0.73$ and 
$\hat{d}_{x,y,z}(1s2s\rightarrow1s2p_{x,y,z})=4.97$ \cite{nist,Wiese2009}.
The XUV and IR fields are linearly polarized along the x-axis and propagate along the z-axis direction, leading to $\varepsilon_x(t) = \varepsilon_{XUV}(t)+\varepsilon_{IR}(t)$ and $\varepsilon_{y,z}(t) =0 $.
The center wavelengths of the XUV and IR fields are set to $58.4 nm$ and $2000nm$, respectively, with their photon energies approximately matching the energy gaps between $1s^2$  and $1s2p$, and between $1s2s$ and $1s2p$, as shown in Figure \ref{fig:1_laser} (a).
Figure \ref{fig:1_laser} (b) illustrates the XUV and IR temporal profiles, defining their time delay.

The last term ($\hat{V}_m$) describes the Zeeman effect with $\hat{V}_m =\hat{\mu}\cdot B_{0} $, where $\hat{\mu}$ is the magnetic dipole moment of the atom and  $B_{0}$ is the external static magnetic field aligned with the z-axis. 
Under the influence of the magnetic field $B_0$, the $1s2p$ state undergoes Zeeman splitting into three sub-states with $m_j=-1,0,$ and $1$, as shown in Figure \ref{fig:1_laser} (a).
Here, we consider only the singlet states of helium for simplicity. 
Since the total electron spin is zero, the coupling between spin and orbital is ignored.
The response of the system, characterized by the time-dependent dipole moment $d_{x,y,z}(t)$, is determined by taking the trace of the product between the time-evolved density matrix and the corresponding dipole transition matrix.
The absorption probability at a given frequency $\omega$ and time delay $\tau$ is given by $S(\omega,\tau)=-2Im[d_x(\omega,\tau)\cdot\varepsilon_x^*(\omega,\tau)+d_y(\omega,\tau)\cdot\varepsilon_y^*(\omega,\tau)+d_z(\omega,\tau)\cdot\varepsilon_z^*(\omega,\tau)]$ \cite{Wu2016}, where $d_{x,y,z}(\omega,\tau)$ is the Fourier transform of $d_{x,y,z}(t,\tau)$, $\varepsilon_{x,y,z}^*(\omega,\tau)$ is the complex conjugate of the Fourier transform of the corresponding total laser fields, respectively. 

\vspace{-0.4cm}

\section{Results and Discussions} \label{sect:dis}
\vspace{-0.4cm}
\subsection{ATAS dressed by a magnetic field} 

\vspace{-0.4cm}

\begin{figure}[hbt]
    \centering
    \includegraphics[width=0.85\linewidth]{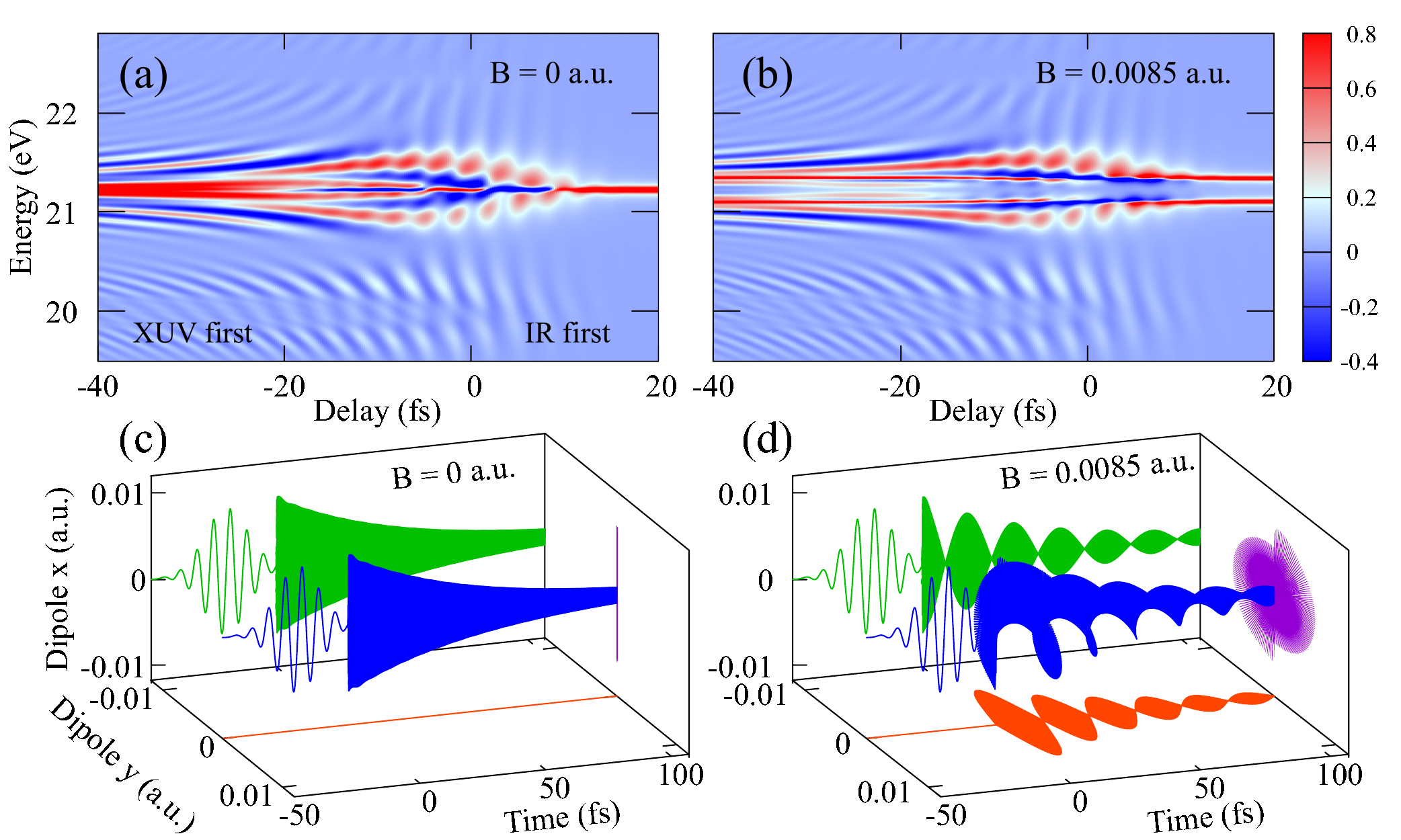}
    \caption{Attosecond transient absorption spectrum as a function of the time delay between the IR and XUV fields without (a) and with (b) an external static magnetic field. Dipole moment as a function of time is shown for a time delay of $20 fs$ for the magnetic field of $B = 0 a.u.$ (c) and $B= 0.0085 a.u.$ (d). }
    \label{fig:2_atas}
\end{figure}

To investigate the influence of the external static magnetic field on ATAS, we simulate the absorption spectrum of helium as a function of the time delay between the IR and XUV pulses.
Figure \ref{fig:2_atas} (a) presents the typical time-resolved attosecond transient absorption spectrum of the Helium atom.
Similar spectra have been observed both experimentally \cite{Chang2013} and theoretically \cite{Wu2016,Xue2022,Chen2013}. 
The minor difference between Figure \ref{fig:2_atas} (a) and previous studies arises from the choice of IR field wavelength. 
Here, an IR field with a central wavelength of $2000 nm$ is used to couple the $1s2p$ and $1s2s$ states. 
In contrast, previous studies typically employed an $800 nm$ IR field, which coupled the $1s2p$ and $1s3s$ states. 
The longer infrared (IR) wavelength reduces the influence of higher-lying excited states, allowing to focus on the interaction between the $1s2p$ and $1s2s$ states.

As depicted in Figure \ref{fig:2_atas} (b), the absorption spectrum clearly exhibits the Zeeman splitting induced by the applied static magnetic field.
This splitting effect is particularly pronounced for positive delays larger than $10 fs$, where the IR pulse has almost completely passed, allowing the absorption spectrum to exclusively reflect XUV-driven transitions between the Zeeman-split sub-states.
Figure \ref{fig:2_atas} (c) and (d) illustrate the dipole moment components along the x-axis (green line) and y-axis (red curve) at a fixed time delay of $20 fs$ with and without the magnetic field, respectively. 
In the absence of the magnetic field [Figure \ref{fig:2_atas} (c)], the dipole moment has only an x-component, induced by the XUV and IR fields. 
However, under the influence of the magnetic field [Figure \ref{fig:2_atas} (d)], a nonzero y-component of the dipole moment emerges, despite the electric fields being linearly polarized along the x-axis.
This transverse response results from the mixing of orbital sub-states induced by Zeeman interaction.
From Figure \ref{fig:2_atas} (d), an oscillatory behavior with the period of $18 fs$ is observed. 
This arises from the interference between the two states, $1s2p_{m_j = -1}$ and $1s2p_{m_j = 1}$, whose energy splitting of $0.23 eV$ precisely matches the observed oscillation frequency.


\begin{figure}[hbt]
    \centering
    \includegraphics[width=0.85\linewidth]{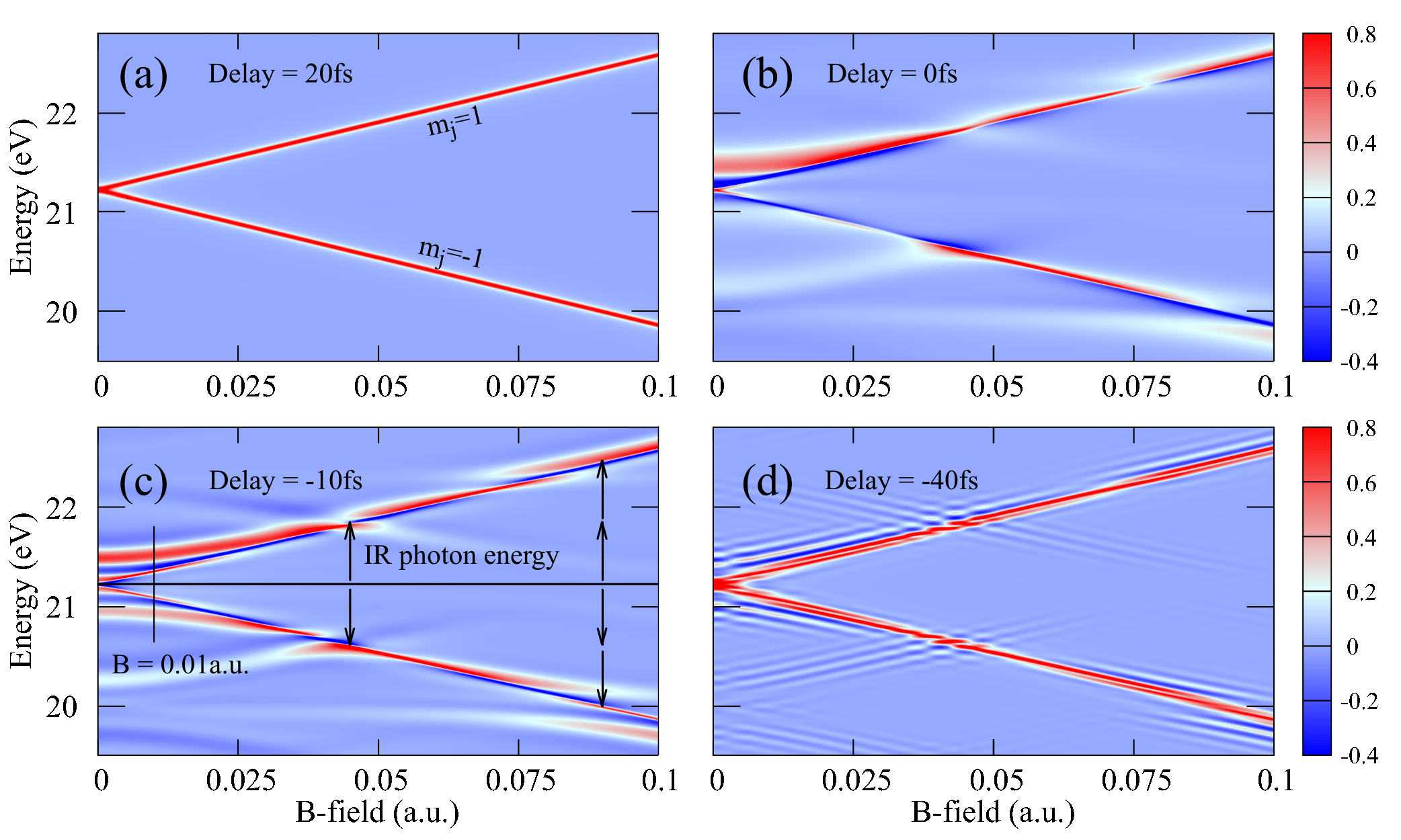}
    \caption{Absorption spectrum as a function of magnetic field strength for different time delays: (a) $20 fs$, (b) $0 fs$, (c) $-10 fs$, and (d) $-40 fs$. The unit of magnetic field strength is expressed in atomic units ($1 a.u. = 2.35 \times 10^5 T$).}
    \label{fig:3_Zeeman}
\end{figure}

Figure \ref{fig:3_Zeeman} presents the Zeeman splitting as a function of magnetic field strength, offering an alternative perspective on the absorption probability of $m_j=-1$ and $m_j=1$ sub-states.
Since the laser is polarized along the x-axis and the magnetic field is aligned along the z-axis, the ground-state wavefunction of Helium can only be excited to the $ m_j = -1 $ or $m_j = 1$ sub-states, due to the dipole transition selection rules. 
Consequently, in Figure \ref{fig:2_atas} (a-d) only two absorption branches, corresponding to $ m_j = -1 $ and $ m_j = 1 $, are observed, rather than all three Zeeman components ($ m_j = -1,0,1 $).
As expected, the Zeeman splitting increases linearly with increasing magnetic field strength \cite{Zeeman1897}.
This rule strictly holds only at certain positive time delays, for example, delay = $20 fs$, as depicted in Figure \ref{fig:3_Zeeman} (a).
For zero and negative delay, Figure \ref{fig:3_Zeeman} (b) and (c), the absorption spectra exhibit more complex features due to the profound involvement of IR-field.
Notably, unique spectral structures can be observed at specific magnetic fields ($B=0.045 a.u.$ and $B=0.090 a.u.$), where the Zeeman-split sub-states strongly resonant with the IR field.
The most significant feature is the distinct absorption asymmetry observed in the branches corresponding to $m_j = -1$ and $m_j = 1$, which will be further discussed in the following section.
Figure \ref{fig:3_Zeeman} (d) shows the result for the time delay of $-40 fs$, where the XUV pulse precedes the IR pulse, with almost no overlap between the two pulses.
The overall structure is similar to that of Figure \ref{fig:3_Zeeman} (a), while the presence of small fringes indicates a weak modification induced by the later-arrived IR pulse.

\vspace{-0.8cm}

\subsection{Asymmetry factor of absorption probability}

\vspace{-0.4cm}

\begin{figure}[hbt]
    \centering
    \includegraphics[width=0.85\linewidth]{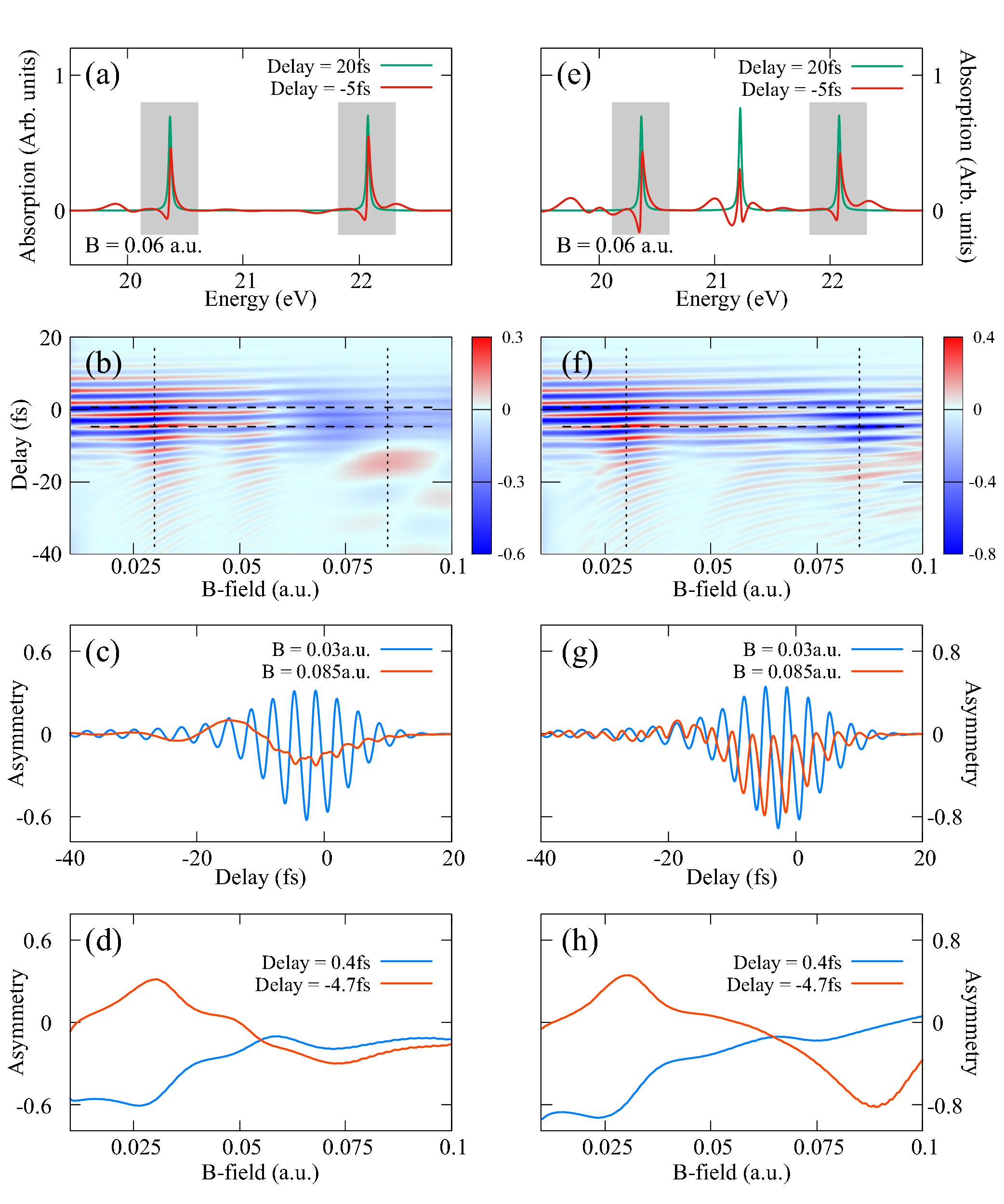}
     \caption{(a) Absorption spectrum highlighting the integration regions for $m_j = \pm1$ (shaded). (b) 2D map of the asymmetry factor as a function of time delay and magnetic field strength, with cross-sections for (c) selected B-values and (d) delay-dependent trends. (e-h) the same as (a-d) but for the case of exciting all the three sub-states of $1s2p$. }
    \label{fig:4_OD_asy}
\end{figure}

We now investigate the asymmetry in absorption probability associated with different orbital magnetic quantum numbers ($m_j = \pm1$).
Since $S(\omega,\tau,B_0)$ represents the absorption probability per unit frequency, the absorption probability is thus obtained by integrating $S(\omega,\tau,B_0)$ over the relevant frequency range. 
Specifically, the absorption probability for $m_j=\pm1$ is calculated using the formula, $I_{m_j}(\tau,B_0)=\int_a^bS(\omega,\tau,B)d\omega$, where $I_{m_j}(\tau,B_0)$ represents the absorption probability, and $a,b$ define the integration limits corresponding to the relevant frequency range.
The two shaded areas in Figure \ref{fig:4_OD_asy} (a) show the integral regions corresponding to $m_j = -1$ (left rectangle) and  $m_j = 1$ (right rectangle), respectively. 
The central position of the integration region, $(a+b)/2$, is determined by Zeeman splitting to ensure that it consistently encompasses the primary absorption peak.
The asymmetry factor for absorption probability is defined as $\frac{I_{m_j=-1}-I_{m_j=1}}{I_{m_j=-1}+I_{m_j=1}}$.
Based on this definition, there is absorption asymmetry even when the magnetic field strength is less than $0.01 a.u.$  as indicated by the vertical line in Figure \ref{fig:3_Zeeman} (c). 
However, this asymmetry arises from the Stark effect induced by the electric field rather than the real asymmetry between the sub-states of $m_j = -1$ and $m_j=1$ caused by Zeeman splitting. 
To eliminate these false asymmetries, the magnetic field strength is set as to start from $0.01 a.u.$ in subsequent analysis.

To provide a comprehensive understanding of how magnetic field strength and time delay influence the asymmetrical factor, we performed a two-dimensional parameter scan, as shown in Figure \ref{fig:4_OD_asy} (b). 
The asymmetry factors as a function of time delay indicated by the two dotted vertical lines in Figure \ref{fig:4_OD_asy} (b), are displayed in Figure \ref{fig:4_OD_asy} (c).
For a relatively weak field [blue curve in Figure \ref{fig:4_OD_asy} (c)], the asymmetry factor exhibits an oscillatory structure with a period of $3.3 fs$, which is half the period of the IR field with central wavelength $2000 nm$.
This indicates that the asymmetry factor is insensitive to the field's orientation and instead depends on its intensity.
While, for a higher magnetic field strength, as illustrated by the red curve in Figure \ref{fig:4_OD_asy} (c), the modulation of the asymmetry becomes significantly weaker, and the oscillatory structure behavior is suppressed.
This trend is further confirmed in Figure \ref{fig:4_OD_asy} (d), which presents the asymmetry factors as a function of magnetic field strength, indicated by the two dashed horizontal lines in Figure \ref{fig:4_OD_asy} (b).
For both time delays (blue and red curves), the asymmetry modulation decreases as the magnetic field strength increases.

Figure \ref{fig:4_OD_asy} (b-d) demonstrate that oscillatory behavior and strong modulation only occur in regions where the magnetic field strength is below $0.05 a.u.$. 
At $0.05 a.u.$, the Zeeman splitting between the $1s2p_{m_j=-1}$ and $1s2p_{m_j=1}$ sub-states induced by the magnetic field approximately equals to twice of the IR photon energy ($1.24 eV$).
There is no single IR photon interaction, as a single-photon process between the $1s2p_{m_j=-1}$ and $1s2p_{m_j=1}$ sub-states is forbidden by the $\Delta L =0$ selection rule.
This suggests that when the Zeeman splitting is smaller than this energy, interference occurs between the coupling pathways $1s^2-1s2p_{m_j=-1}-1s2s$ and $1s^2-1s2p_{m_j=1}-1s2s$, resulting in oscillatory behavior and asymmetry modulation. 
As the magnetic field strength increases beyond this threshold, the Zeeman splitting exceeds two IR photon energy, diminishing the interference effect, and consequently reducing the asymmetry modulation.

\vspace{-0.8cm}

\subsection{The influence of the $m_j=0$ sub-state}

\vspace{-0.4cm}

So far, our study has primarily focused on the case where the laser fields are polarized along the x-axis, selectively exciting the ground state into the orbital sub-states of $m_j = -1$ and $m_j = 1$. 
To achieve simultaneous excitation into all three sub-states ($m_j=\pm1,0$), it is sufficient to adjust the angle between the laser polarization and the magnetic field direction so that it deviates from $90^\circ$, ensuring that the laser fields has a nonzero projection along the z-axis.
The polarization state of the electric fields remain unchanged as linear polarization.
Given that polarization angle adjustment is experimentally straightforward, it is worthwhile to examine the influence of the $m_j = 0$ sub-state on the asymmetry modulation of $m_j = -1$ and $m_j = 1$.

Figure \ref{fig:4_OD_asy} (e-h) present the results for the case where all three sub-states ($1s2p_{m_j=\pm1,0}$) are excited equally. 
Here, "equally" means that the transition probabilities from the ground state to each of the three excited sub-states ($1s2p_{m_j=\pm1,0}$)  are identical, promising the strongest possible interference among them.
The requirement for identical excitation is that the angle between the laser polarization and the magnetic field must be $ 54.7^\circ $, the magic angle - an important concept in magnetic resonance imaging \cite{Bydder2007}.
As shown by the green curve in Figure \ref{fig:4_OD_asy} (e), the absorption peaks corresponding to $m_j=-1,0,1$ exhibit nearly identical heights.
A comparison between Figure \ref{fig:4_OD_asy} (f-h) and Figure \ref{fig:4_OD_asy} (b-d) reveals that the most significant difference occurs around a magnetic field strength of approximately $ 0.09  a.u.$. 
As shown by the red curve in Figure \ref{fig:4_OD_asy} (g), the asymmetry factor exhibits an oscillatory structure, and the amplitude of the asymmetry modulation is also enhanced - features that are absent in the $ 90^\circ $ case (Figure \ref{fig:4_OD_asy} (b-d)).
The red curve in Figure \ref{fig:4_OD_asy} (h) further confirms that difference.
This behavior is understandable, as a magnetic field strength of $ 0.09 $ a.u. leads to the energy splitting between the $1s2p_{m_j=\pm1}$ and $1s2p_{m_j=0}$ sub-states roughly equals to the energy of two IR photons.
Consequently, the two coupling pathways ($1s^2-1s2p_{m_j=-1,1}-1s2s$) can interfere with the third coupling pathway ($1s^2-1s2p_{m_j=0}-1s2s$) through IR-induced transitions, leading to the emergence of oscillatory structures and strong asymmetry modulation. 
For magnetic field strengths below $ 0.05  a.u.$, the results for $ 54.7^\circ $ and $ 90^\circ $ are highly similar, indicating that the influence of the $ m_j = 0 $ state is not pronounced in this regime.
This coupling-pathway-interference explanation is further confirmed by the population asymmetry results for angles of $90^\circ$ and $54.7^\circ$, as shown by Fig. S1 in the Supplementary Material. 
Together, Figures \ref{fig:4_OD_asy} (a-h) demonstrate that by appropriately selecting the time delay, the magnetic field strength, and the angle between laser polarization and magnetic field direction, the desired asymmetry of orbitals with $m_j = -1$ or $m_j = 1$ can be achieved.
This tunability highlights the feasibility of manipulating OAM states through external field parameters.

\vspace{-0.7cm}
\subsection{Experimental Feasibility and Future Applications}

\vspace{-0.35cm}

The maximum magnetic field strength used in this study is approximately $0.1 a.u.$ ($2.35\times10^4 T$). 
However, this value is not a strict requirement.
As shown in Figure \ref{fig:4_OD_asy} (b), even for relatively low magnetic field strengths considered, such as $0.01 a.u.$ ($2.35\times10^3 T$), the IR-$2\omega$ oscillation remains observable. 
This indicates that the OAM polarization can be effectively achieved even at lower magnetic field strengths.
Moreover, laboratory experiments have successfully demonstrated the generation of kilo-tesla magnetic strengths, as reported by various research groups \cite{Fujioka2013,Nakamura2018}.
Research utilizing or generating strong magnetic fields has also overcome many limitations and yielded remarkable results \cite{Martin2023,Heras2023,Sederberg2020}.
Given these advancements, we anticipate that the application of strong magnetic fields in experimental settings will become increasingly feasible in the near future.

The presented methodology paves the way for many important advancements.
For instance, it could enable the precise manipulation of chemical reactions, which are heavily influenced by the spatial distribution of electron orbitals. 
Additionally, it could facilitate the modification of the magnetic properties of atoms and molecules, as these properties are governed by orbital magnetic quantum numbers.
Such control mechanisms could also unlock new techniques for tailoring magnetically sensitive materials, opening avenues for advancements in nanotechnology and photonics.
Furthermore, this method could pave the way for the development of a novel type of laser that emits circularly polarized light, as the selection rule for such lasers depends on modifying the orbital magnetic quantum number.

\section{Conclusion}\label{sect:conc}

We have presented a novel approach for controlling the OAM polarization of electrons in neutral atomic systems by integrating the Zeeman effect with attosecond transient absorption spectroscopy. 
Using density matrix simulations, we demonstrate that OAM polarization can be precisely tuned by adjusting the time delay between IR and XUV fields, the static magnetic field strength, as well as the angle between the laser polarization and magnetic field direction.
This method not only deepens our understanding of ultrafast electron dynamics but also introduces a new degree of control over quantum state engineering. The implications of this methodology extend across diverse fields.
Future work should focus on experimental realizations of this technique, particularly at lower magnetic field strengths that align with current laboratory capabilities. 
Investigating its applicability in more complex atomic and molecular systems could further validate the robustness of this method and expand its technological potential. 
As high-field laser physics continues to evolve, these findings could serve as a foundation for next-generation quantum technologies and ultrafast control of electronic states.

\bibliography{ref_bib}



\newpage

\begin{center}
\textbf{\Large{Supplementary Material}}    
\end{center}

\renewcommand{\thefigure}{S\arabic{figure}}
\setcounter{figure}{0}  
\setcounter{page}{1}

\subsection{Cartesian basis and the matrix transformation}

In this work, Cartesian basis refers to the selection of the $1s2p$ sub-states, namely $1s2p_x$, $1s2p_y$, and $1s2p_z$. 
Its advantage is that the electric transition dipole moments are clear and straightforward.
The corresponding angular momentum matrix is non-diagonal since the Cartesian basis are not eigen states of the angular momentum operator.
Specifically, the total Hamiltonian,$\hat{H}=\hat{H}_0+\hat{V}_e+\hat{V}_m$, in Cartesian basis is as follows,
 \begin{equation}
 \label{equ:matrix} 
  \begin{bmatrix}
          E_{1s^2}&0&0&0&0\\
          0&E_{1s2s}&0&0&0\\
          0&0&E_{1s2p_x}&0&0\\
          0&0&0&E_{1s2p_y}&0\\
          0&0&0&0&E_{1s2p_z}
  \end{bmatrix}
  + \begin{bmatrix}
          0&0&d_x^1 \epsilon_x&d_y^1 \epsilon_y&d_z^1 \epsilon_z\\
          0&0&d_x^2 \epsilon_x&d_y^2 \epsilon_y&d_z^2 \epsilon_z\\
          d_x^1 \epsilon_x&d_x^2 \epsilon_x&0&0&0\\
          d_x^1 \epsilon_y&d_x^2 \epsilon_y&0&0&0\\
          d_x^1 \epsilon_z&d_x^2 \epsilon_z&0&0&0
  \end{bmatrix}
  + \begin{bmatrix}
          0&0&0&0&0\\
          0&0&0&0&0\\
          0&0&0&i\mu B_{0}&0\\
          0&0&-i\mu B_{0}&0&0\\
          0&0&0&0&0
  \end{bmatrix} 
  \end{equation}
  where, $d_{x,y,z}^1=d_{x,y,y}(1s^2\rightarrow1s2p)$ and $d_{x,y,z}^2=d_{x,y,y}(1s2s\rightarrow1s2p)$. 
  $\mu$ equals to $m_j\cdot g_j \cdot u_B$, where $g_j$ is the Landé $g$-factor ($=1$ for $1s2p$), and $u_B$ is the Bohr magneton.

Alternatively, one can use the energy basis, in which the sub-states of $1s2p$ are set as $1s2p_{m_j=-1}$, $1s2p_{m_j=0}$, and $1s2p_{m_j=1}$. 
Those two basis give the same results, and they can be transformed between each other, following the relation of $\hat{\rho}_{mj=\pm,0}=P^{-1}\hat{\rho}_{x,y,z}P$.
The matrix of $P$ can be obtained from the diagonalization of the angular momentum matrix.
The dipole relation between those two basis are also straightforward, e.g. $d_{m_j=-1}=(d_x+id_y)/2$, $d_{m_j=1}=(d_x-id_y)/2$, and $d_{m_j=0}=d_z$.
Note: The dipole relation between those two basis are also straightforward, e.g. $d_{m_j=-1}=(d_x+id_y)/\sqrt{2}$, $d_{m_j=1}=(d_x-id_y)/\sqrt{2}$, and $d_{m_j=0}=d_z$.

\subsection{Population asymmetry}

\begin{figure}[hbt]
    \centering
    \includegraphics[width=0.90\linewidth]{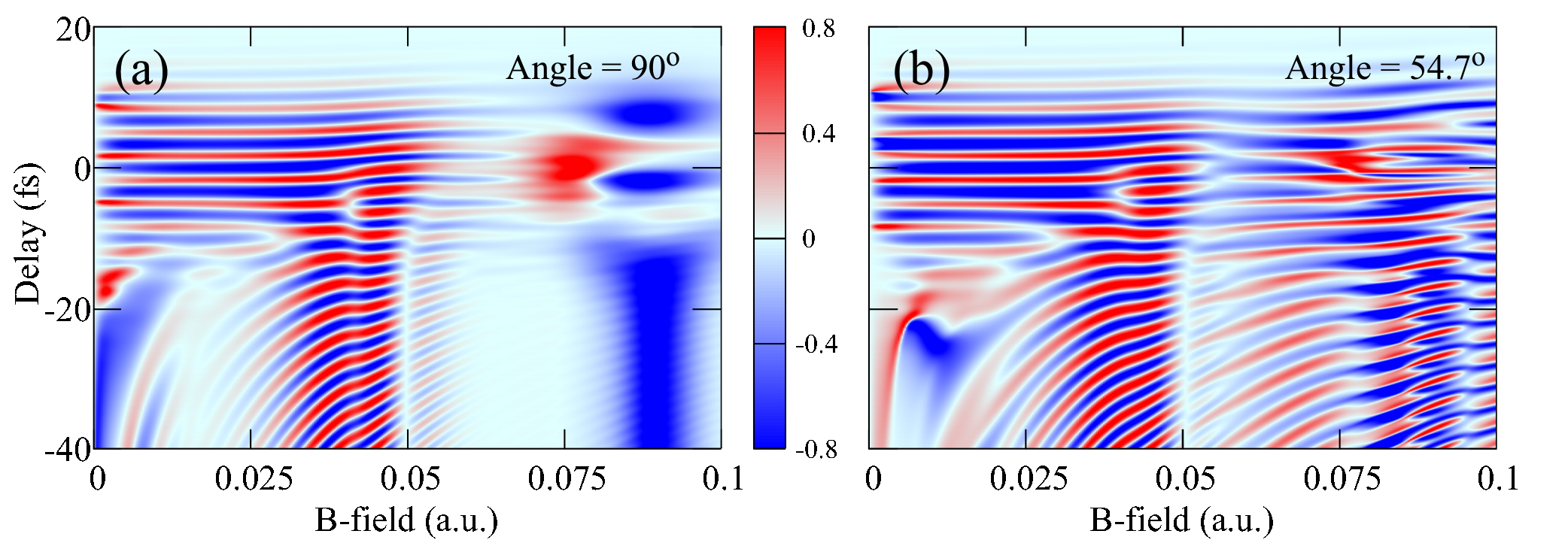}
    \caption{
    The population asymmetry factor as a function of magnetic field strength and time delay for the angle of $90^\circ$ (a) and $54.7^\circ$ (b).}
    \label{fig:5_pop_asy}
\end{figure}

Figure \ref{fig:5_pop_asy} shows the population asymmetry factor of for $m_j = -1$ and $m_j = 1$ after the interaction between the atom and the IR+XUV fields when the angle between the laser polarization and magnetic field direction is $90^\circ$ (a) or $54.7^\circ$ (b). 
The overall structure of the population asymmetry in Figure \ref{fig:5_pop_asy} (a) and (b) closely resembles the absorption asymmetry in Figure 4 (b) and (f), respectively.
This similarity arises from the fact that the dipole moment is determined by both the diagonal elements (state populations) and the non-diagonal elements (coherence terms) of the dipole matrix.
When the coherence is weak, the population plays the dominant role in the absorption process, leading to similar structures in both population and absorption asymmetry.
Thus, the regions where these similarities occur indicate that the population dominates the absorption process.

The compare between Figure \ref{fig:5_pop_asy} (a) and (b) further confirms the coupling-pathway-interference explanation.
One can see that for the case of exciting two states ($90^\circ$), oscillatory structures only appear in the range of $B<0.05 a.u.$.
While, when three states are excited simultaneously ($54.7^\circ$), oscillatory structures can be observed in both ($B<0.05 a.u.$ and $B=0.09 a.u.$) ranges.

\end{document}